# Leveraging high-performance head-only gradients to enable ~2× higher multiband acceleration in SMS-EPI and SMS-EPTI

Yimeng Lin[1,2], Zachary Shah[1,2], Mehdi Khalighi[2], Zimu Huo[2], Mengze Gao[2], Hua Wu[7], Baolian Yang[4], Qiang Liu[5,6], Erpeng Dai[5], Arthur Chakwizira[5], Lipeng Ning[5], Carl-Fredrik Westin[5], Yogesh Rathi[5], Jonathan R. Polimeni[2], Nan Wang[3], Adam B. Kerr[1,7], Kawin Setsompop[1,2,7]

1. Department of Electrical Engineering, Stanford University, CA, USA
2. Department of Radiology, Stanford University, CA, USA
3. Department of Radiology, University of California San Francisco, CA, USA
4. MR, GE Healthcare, Waukesha, Wisconsin, USA
5. Department of Psychiatry, Brigham and Women's Hospital, Harvard Medical School, Massachusetts, USA
6. College of Engineering, Northeastern University, Massachusetts, USA
7. Cognitive and Neurobiological Imaging Center, Stanford University, California, USA

## Synopsis

**Motivation:** High-performance gradients enable efficient 3D k-space traversal in SMS-EPI and SMS-EPTI acquisitions with denser $k_y$–$k_z$ sampling to support higher MB accelerations.

**Goals:** Improve SMS-EPI and SMS-EPTI acceleration to achieve higher temporal resolution for whole-brain fMRI.

**Approach:** Fast k-space traversal with denser $k_y$–$k_z$ sampling is achieved via PNS-optimized gradient waveform design while high-quality reconstruction is obtained via field-correcting GRAPPA (FCG).

**Results:** Approximately 2× faster k-space traversal on MAGNUS 3T and Impulse 7T when compared to high-performance whole-body systems enables ~2× higher MB. This facilitates high-quality 2.5 mm whole-brain rs-fMRI at 7T at MB-28 and TR = 123 ms.

## Impact

Optimized utility of high-performance head-only systems enables 3D k-space traversal with denser $k_y$–$k_z$ sampling in SMS-EPI and SMS-EPTI to support ~2× increase in MB accelerations, enabling higher spatiotemporal resolution fMRI.

## Introduction

Emerging high-performance head-only systems facilitate substantially faster echo-planar imaging (EPI) encoding,[1,2] while advances in field-error correction allow high-fidelity reconstruction of these fast acquisitions.[3,4] Shorter echo spacing (ESP) is commonly used to shorten the readout and reduce distortion. On the other hand, it can also provide an opportunity to acquire more $k_y$–$k_z$ encoding within the same readout duration (at the same distortion level) to support higher multiband (MB) acceleration.[5,6] Here, we use peripheral nerve stimulation (PNS)-optimized gradient waveform design and field-correcting reconstruction to achieve ESPs on head-only 3T and 7T systems that are approximately half those of high-performance whole-body systems. This facilitates 2× additional $k_y$–$k_z$ encodings across the same readout window to improve simultaneous-multislice (SMS) EPI acceleration and the spatiotemporal trade-off for whole-brain fMRI. We also extend this strategy to single-shot echo planar time-resolved imaging (EPTI).

## Methods

**Enabling short-ESP acquisition.** To exploit the high gradient amplitude and slew rate of head-only systems within PNS limits, we optimized EPI readout waveforms on MAGNUS 3T using linear programming with hardware constraints and a convolution-based PNS model.[7,8] This formulation allows flexible waveform shapes without prescribing a specific gradient pulse profile. The resulting waveform's slew rate starts high, then decreases as the predicted PNS response increases (Figure 1a), similar to recently proposed PNS-aware pulse shapes[9].

Fast readouts also increase ramp sampling and can accentuate spatiotemporally varying eddy-current phase errors. We therefore acquired single-band dual-polarity EPI calibration data to train field-correcting GRAPPA (FCG)[3] kernels. Figure 1a compares phantom reconstructions using linear ghost correction (LGC), dual-polarity GRAPPA (DPG)[4], and FCG, which illustrates the importance of FCG in achieving clean reconstructions of these data.

**Using faster encoding for higher MB acceleration.** With 2× shorter ESP, twice as many readout lines can be acquired within the original readout window (Figure 1b). A simple approach to employ the extra lines is to halve the $\Delta k_y$ step along the EPI-readout to double the $\mathrm{FOV}_y$, while maintaining nominal spatial resolution. With CAIPI-shifting[5,10], this spreads aliased voxels farther apart across $2\mathrm{FOV}_y$, improving the coil-sensitivity encoding available for slice separation. Figure 1c illustrates an MB6 pattern that retains the $k_z$-state cycle while increasing aliasing voxel separation across slices. Other allocations between $k_y$ sampling and $k_z$-state diversity are possible, where in this work, the CAIPI-shift factor is optimized for each MB case to minimize *g*-factor.

**Evaluation.** SMS-EPI acquired with i) standard-ESP and ii) short-ESP at $2\mathrm{FOV}_y$ were compared using analytical *g*-factor, with reconstruction performed via FCG field-correction followed by LeakBlock slice-GRAPPA.[11] For single-shot SMS-EPTI,[12-14] the additional echoes were used to provide denser $k_y$–time sampling, and its benefit was evaluated at MB2 and MB4 via *g*-factor analysis of FCG-corrected subspace reconstruction. Pilot resting-state fMRI was assessed using posterior cingulate cortex (PCC)-seeded functional connectivity, independent component analysis (ICA)[15,16], and temporal frequency spectra showing physiological noise components.

## Results

Across the tested 2.5–3.14 mm EPI protocols, ESP ranges from 0.248 to 0.290 ms on head-only MAGNUS 3T and Impulse 7T systems, which are ~2× shorter than typically achieved on high-performance whole-body systems (0.480–0.560 ms). Figures 2 and 3 demonstrate the benefit of employing this faster encoding for denser $k_y$–$k_z$ sampling in SMS-EPI: on MAGNUS 3T, high-quality imaging can be obtained at MB14, with *g*-factor only slightly higher than conventional MB6 acquisition; on Impulse 7T, it enables high-quality MB28 acquisition with *g*-factor similar to that of the conventional MB16 case, which facilitates high-quality 2.5 mm isotropic whole-brain imaging at a TR of 123 ms.

Figure 4 extends our approach to 3T single-shot SMS-EPTI at 2.5 mm isotropic resolution, where the denser *k*–*t* sampling facilitates an increase in MB acceleration of this time-resolved distortion-free imaging from MB2 to MB4.

Figure 5 shows the results from a pilot 2.5 mm whole-brain rs-fMRI at 7T at TR = 123 ms during a 4-minute scan. The data shows recognizable PCC-seeded functional connectivity and dorsal-attention, somatomotor, visual, and frontoparietal ICA networks. Image-derived components exhibited putative respiratory fluctuations and cardiac-related harmonic peaks near 1.87 and 2.80 Hz that can be captured with this rapid imaging.

## Conclusion

The use of faster k-space traversal on head-only systems to provide dense $k_y$–$k_z$ sampling and field-correcting-GRAPPA (FCG) to achieve high-quality reconstruction has facilitated SMS-EPI and SMS-EPTI acquisitions at ~2-fold higher MB factors. For SMS-EPI, this has enabled whole-brain fMRI at very high temporal sampling. For single-shot SMS-EPTI, it has enabled high-quality MB4 imaging, and allows this distortion-free multi-echo imaging to be performed at a temporal sampling rate approaching that of conventional SMS-EPI.

## Acknowledgement

This work was supported in part by the National Institutes of Health (grants R01-MH116173; R21-EB038677; R01-HD114719; R01-EB033206; and U24-NS129893) and was made possible by the resources provided by NIH Shared Instrumentation Grant S10-OD036337 and by the C-ShaRP Instrumentation & Enhancement Program at Stanford University.

## Figures

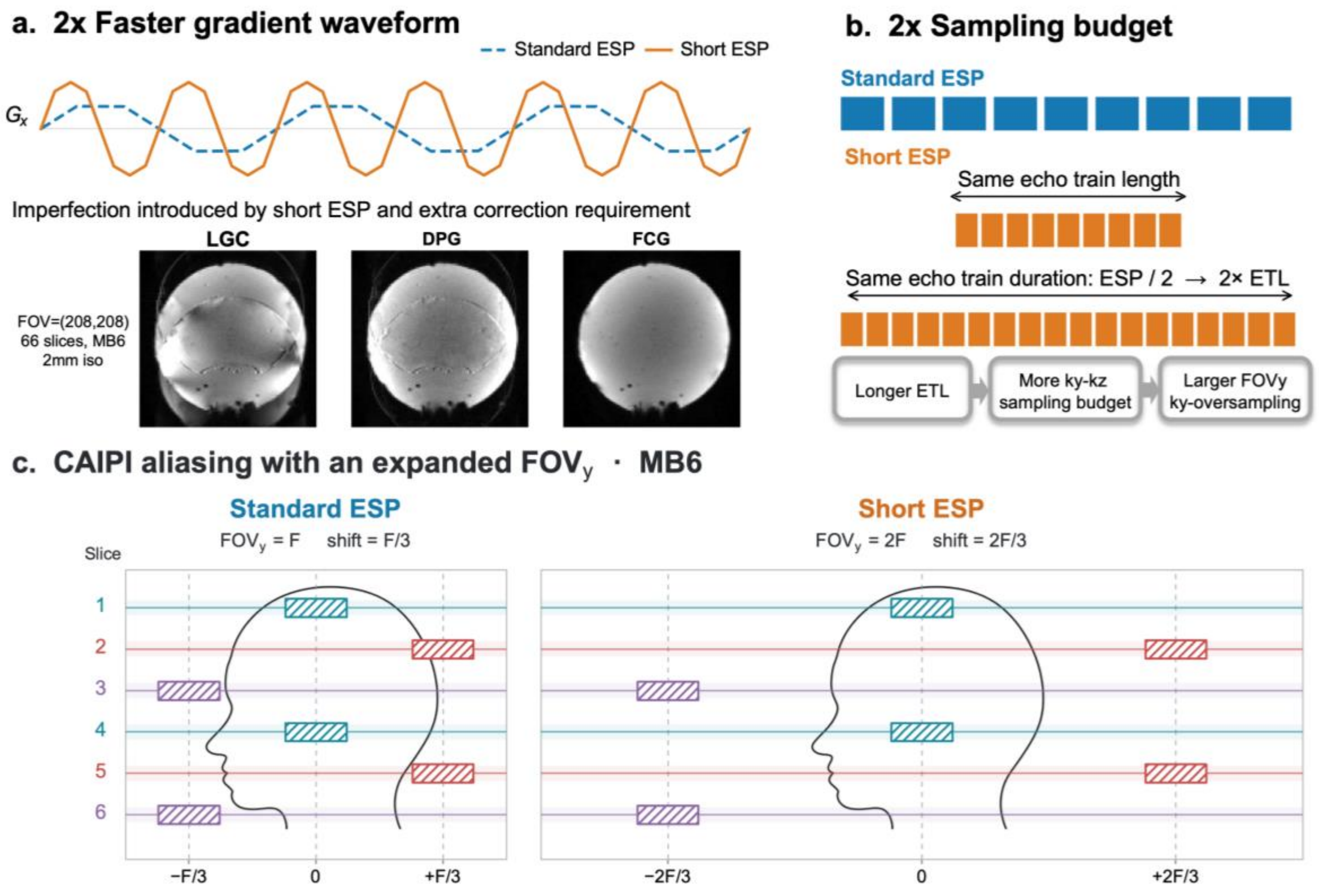

**Figure 1.** (a) Standard/short-ESP gradient designs and phantom reconstructions using LGC, DPG, or FCG. (b) Nearly halved ESP doubles the echo budget within the original readout window. (c) An illustrative MB6 CAIPI allocation: $k_y$ oversampling doubles FOV$_y$ and increases apparent interslice separation while retaining the $k_z$-state cycle; other $k_y$–$k_z$ allocations are possible.

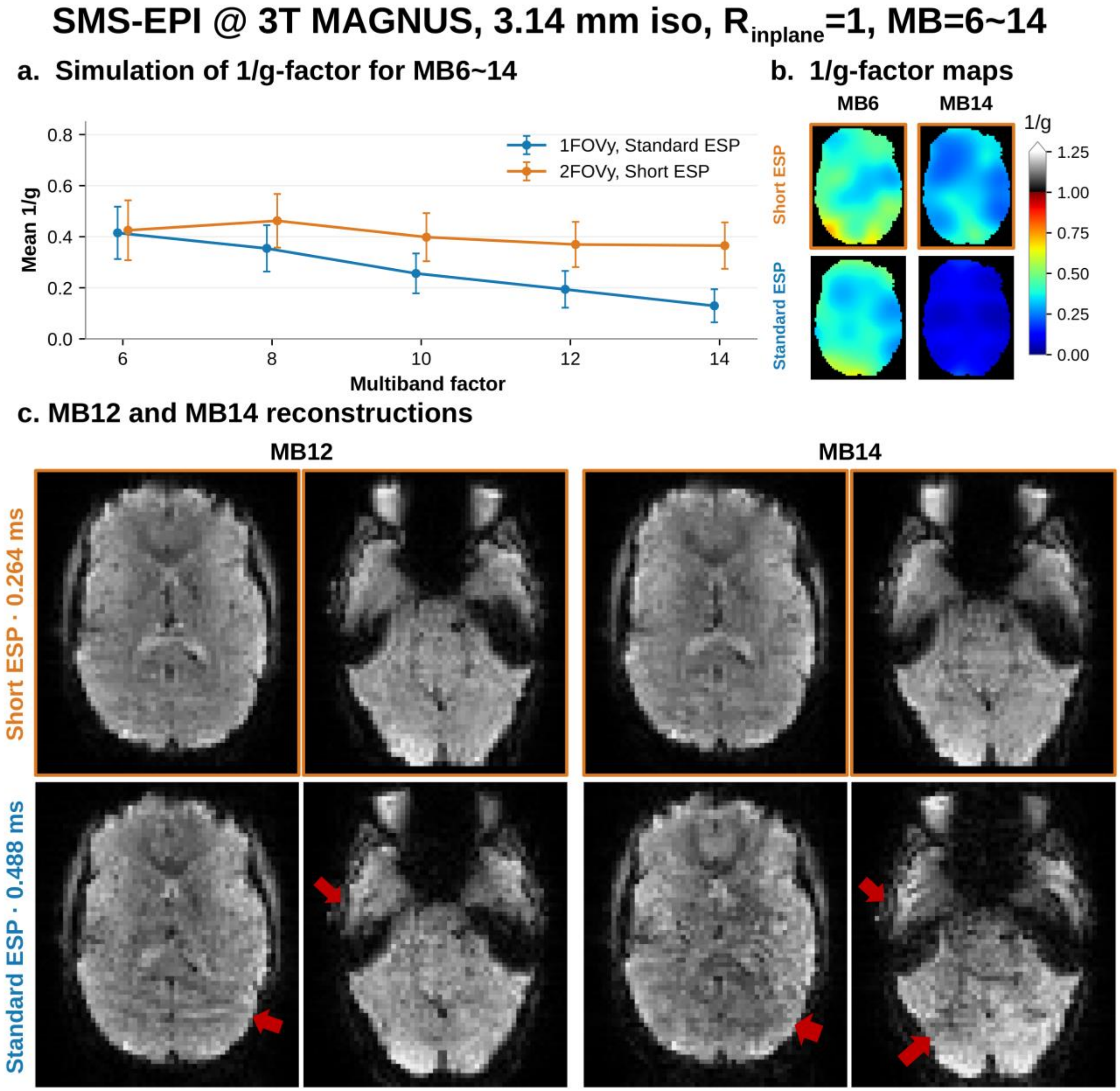

**Figure 2.** 3T MAGNUS SMS-EPI, 3.14 mm isotropic, MB6–14. Standard ESP/1FOV$_y$ (blue) versus short ESP/2FOV$_y$ (orange): (a) analytical mean voxelwise $1/g \pm$ spatial SD. At MB14, mean $1/g$ increases from 0.129 to 0.365. (b) MB6/MB14 $1/g$-factor maps; (c) two slices each at MB12/MB14, reconstructed with FCG plus LeakBlock slice-GRAPPA. Red arrows highlight degraded regions with standard ESP.

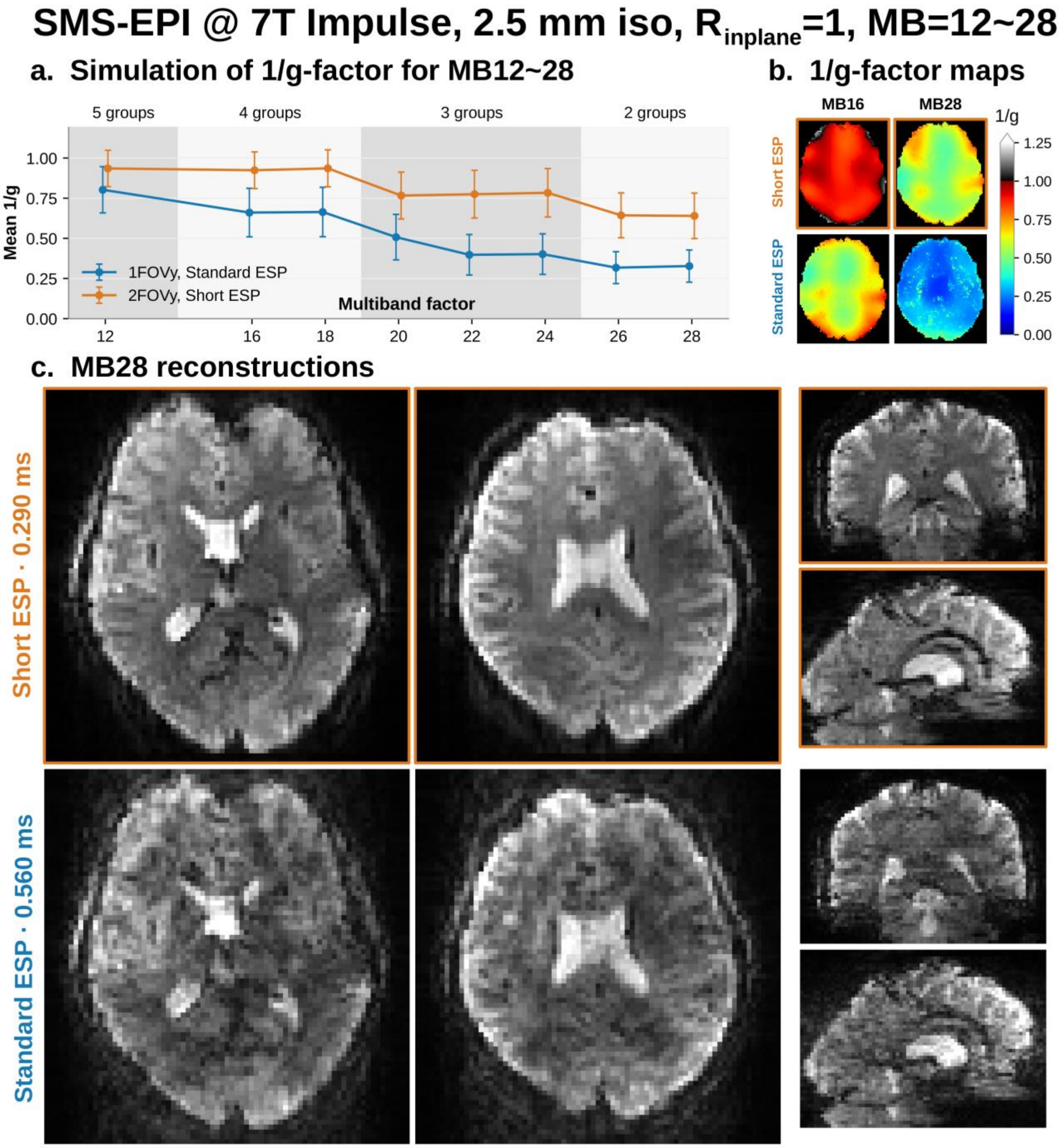

**Figure 3.** 7T Impulse SMS-EPI, 2.5 mm isotropic, MB12–28. (a) Analytical mean voxelwise $1/g$ ± spatial SD; shading indicates slice-group counts. At MB28, mean $1/g$ increases from 0.327 to 0.640. (b) MB16/MB28 $1/g$-factor maps. (c) MB28 FCG plus LeakBlock slice-GRAPPA reconstructions: two axial slices and coronal/sagittal views from one volume.

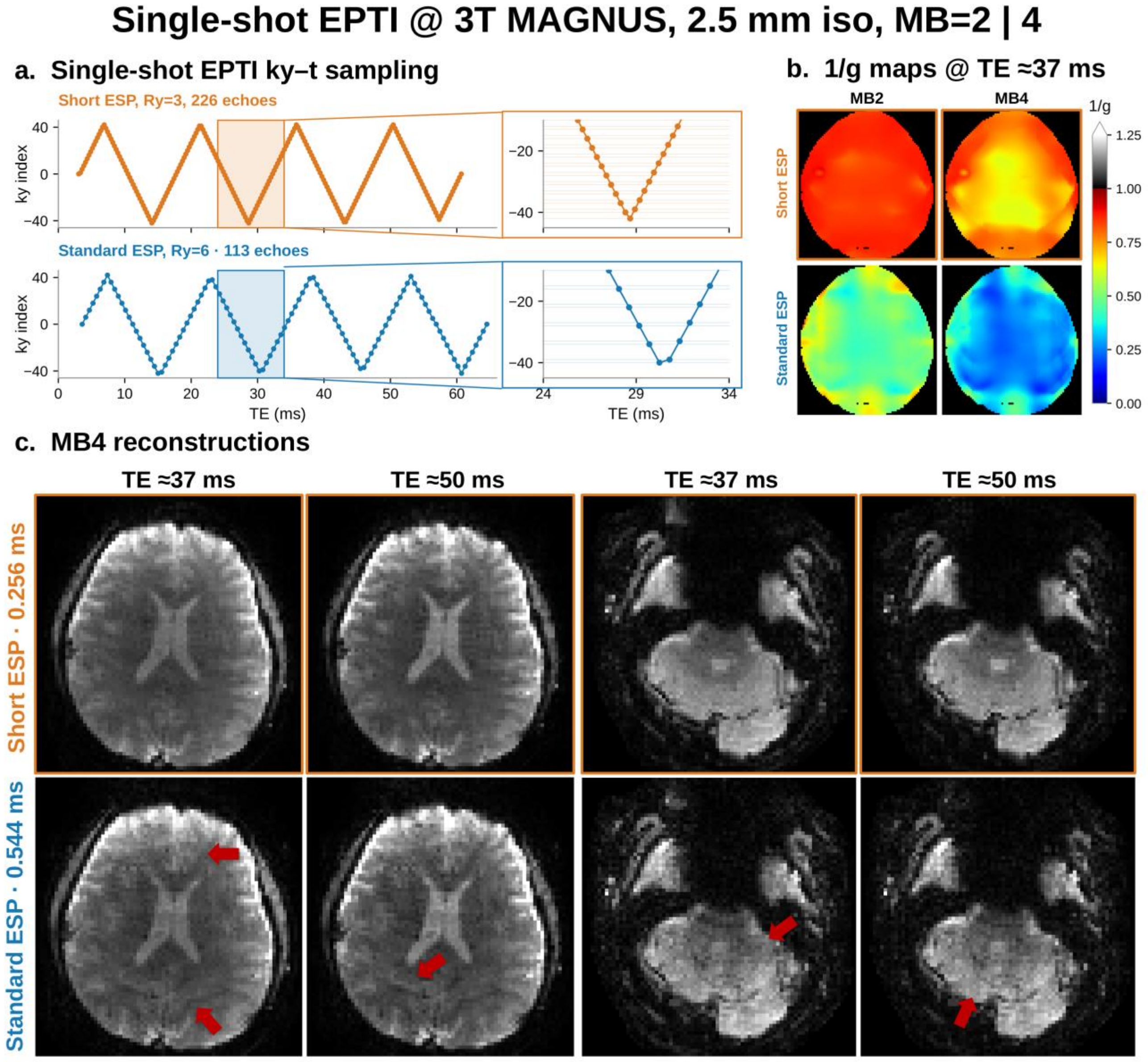

**Figure 4.** 3T MAGNUS single-shot SMS-EPTI, 2.5 mm isotropic. (a) Short ESP doubles echo count (113 to 226) and reduces $R_y$ from 6 to 3; insets enlarge the $k_y$–t sampling. (b) MB2/MB4 unregularized-subspace 1/$g$ maps at TE ≈ 37 ms. (c) Two MB4 slices at TE ≈ 37/50 ms, using FCG-corrected subspace reconstruction with $B_0$ updates and LLR. Orange frames indicate short ESP; red arrows highlight degraded regions with standard ESP.

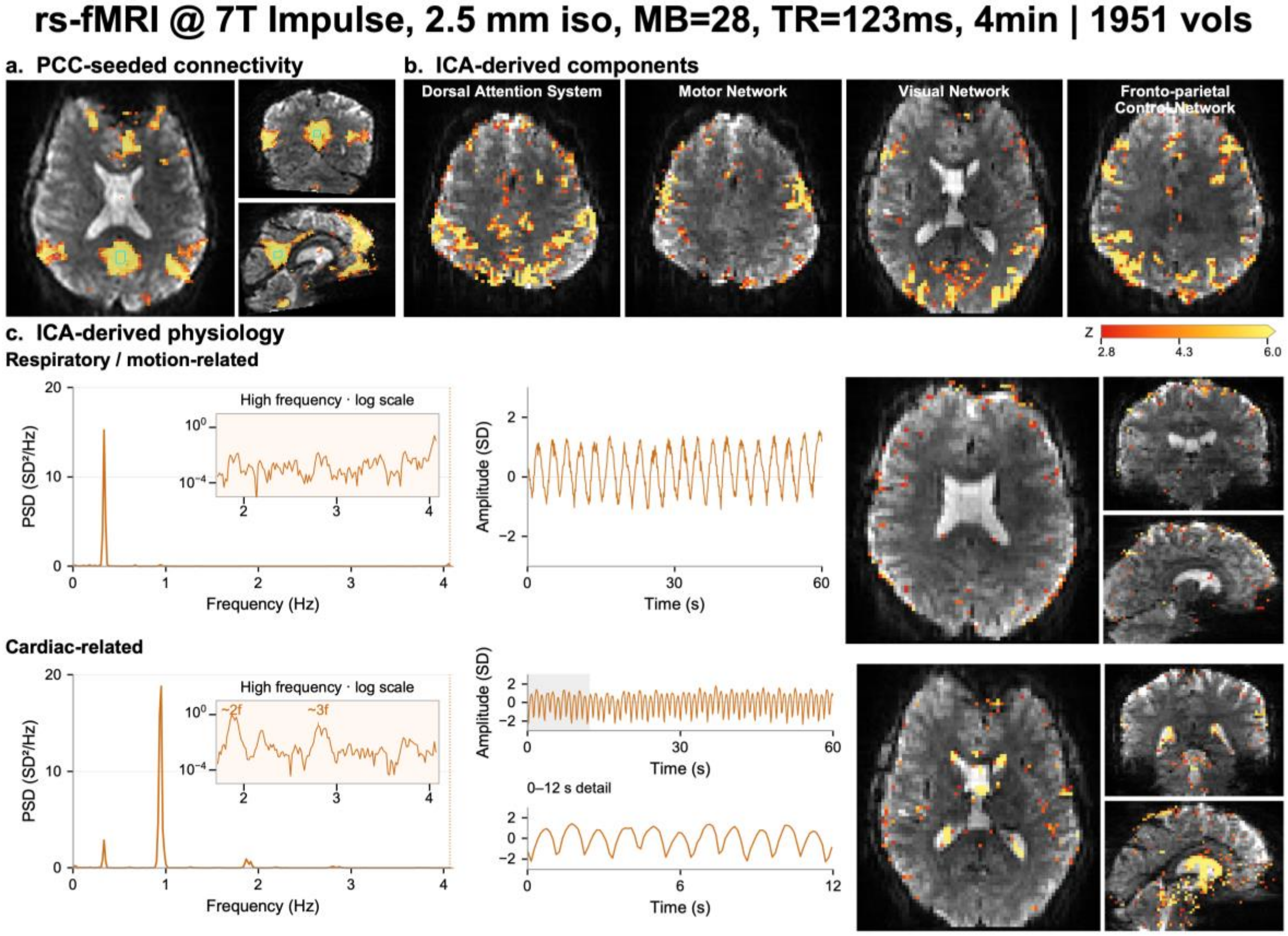

**Figure 5.** Single-subject 7T rs-fMRI, 2.5 mm isotropic. (a) PCC connectivity: from 4-minute MB28/2FOV$_y$ (TR = 123 ms, 1951 retained volumes). Native motion-reference backgrounds. (b) MB28 ICA networks. (c) MB28 respiratory/motion and cardiac components, spatial maps, time courses, and power spectra. $Z > 2.8$.